\documentclass[conference]{IEEEtran}
\IEEEoverridecommandlockouts

\usepackage{cite}                    
\usepackage{amsmath,amssymb,amsfonts}
\usepackage{algorithmic}
\usepackage{graphicx}                
\usepackage{textcomp}
\usepackage{xcolor}
\usepackage{booktabs}                
\usepackage{multirow}                
\usepackage{array}
\usepackage{subcaption}              
\usepackage{url}
\usepackage{balance}                 
\usepackage[hidelinks]{hyperref}     

\def\BibTeX{{\rm B\kern-.05em{\sc i\kern-.025em b}\kern-.08em
    T\kern-.1667em\lower.7ex\hbox{E}\kern-.125emX}}

\newcommand{\R}{\mathrm{ReLU}}
\newcommand{\Rs}{\mathrm{ReLU6}}
\newcommand{\GN}{\mathrm{GN}}
\newcommand{\CS}{\mathrm{CS}}

\begin{document}

\title{3D-EMCAD: Efficient Multi-scale Convolutional Attention Decoding
for Volumetric Brain Tumor Segmentation}

\author{
\IEEEauthorblockN{Iftakhar Hossain}
\IEEEauthorblockA{Department of Biomedical Engineering\\
Bangladesh University of Engineering \& Technology (BUET)\\
Dhaka, Bangladesh\\
iftakhar.bme.buet20@gmail.com}
\and
\IEEEauthorblockN{Israt Jahan Sinthia}
\IEEEauthorblockA{Department of Biomedical Engineering\\
Chittagong University of Engineering and Technology (CUET)\\
Chittagong, Bangladesh\\
sinthiasrt@gmail.com}
\and
\IEEEauthorblockN{Md. Hadiur Rahman Khan\thanks{Corresponding author: hrkhan@eee.buet.ac.bd}}
\IEEEauthorblockA{Department of Electrical and Electronic Engineering\\
Bangladesh University of Engineering \& Technology (BUET)\\
Dhaka, Bangladesh\\
hrkhan@eee.buet.ac.bd}
}

\maketitle
\thispagestyle{empty}
\pagestyle{empty}

\begin{abstract}
Encoder--decoder networks are widely used for volumetric brain tumor
segmentation, yet recent trends favor high-capacity decoders that incur
steep computational costs in 3D at high spatial resolutions. The
Efficient Multi-scale Convolutional Attention Decoder (EMCAD) is a
lightweight decoder that achieves strong accuracy at very low cost, but
it has been validated only in 2D. Here, we extend EMCAD to volumetric
segmentation (3D-EMCAD) by converting all operations to 3D and replacing
standard normalization layers with group normalization to suit the small
batch sizes of volumetric training. To isolate the impact of decoder
choice, we evaluate five decoders: the proposed 3D-EMCAD, a vanilla
SegResNet decoder, an Attention U-Net decoder, UNet++, and a plain U-Net
decoder using an identical 3D SegResNet encoder and training protocol
on BraTS 2020 ($369$ cases, three-fold cross-validation). The five
decoders yield statistically indistinguishable performance: their mean
Dice scores span only $0.0071$, less than the fold-to-fold variation of
most models, and a paired Wilcoxon signed-rank test shows no significant
difference between 3D-EMCAD and three of the four baselines. Importantly,
3D-EMCAD achieves this shared accuracy with the fewest decoder parameters
($1.15$~M) and operations ($8.6$~G), requiring $7.3\times$ fewer
operations than UNet++, which scores no higher. These findings
demonstrate that decoder capacity beyond a modest budget offers no
measurable benefit for volumetric brain tumor segmentation.
\end{abstract}


\section{Introduction}
Gliomas are among the most common and most aggressive primary brain
tumors, and their clinical management depends on accurate delineation of
the tumor and its sub-regions from multi-parametric magnetic resonance
imaging (MRI). Manual delineation of volumetric MRI scans is laborious
and subject to high inter-rater variability, motivating the development of automatic
segmentation \cite{menze2014brats, bakas2017advancing}. Encoder--decoder
networks have emerged as the standard solution: an encoder compresses the
volume into a hierarchy of semantic features, while a decoder
reconstructs a full-resolution segmentation from
it \cite{ronneberger2015unet, cicek20163dunet}.

Architectural innovation, however, has been unevenly distributed between
these two components. Encoders have been extensively redesigned, whereas
decoders have primarily accumulated capacity through nested skip
pathways, dense connections, and attention modules, under the assumption
that larger decoders inherently improve segmentation quality. This
assumption is particularly costly for volumetric segmentation, where the
decoder operates across high spatial resolutions, the cost of a dense
$k \times k \times k$ convolution grows with $k^3$, and activation memory
grows rapidly with the volume. Consequently, the decoder can consume a
disproportionate share of the total computational budget, straining
hardware resources in typical clinical environments.

Several established decoders illustrate this trend towards greater
capacity: UNet++ \cite{zhou2018unetpp} introduces densely nested skip
pathways, Attention U-Net \cite{oktay2018attention} gates skip
connections with learned attention coefficients, and
SegResNet \cite{myronenko2019segresnet} employs a residual decoder that
is highly effective on brain tumor data, while self-configuring
pipelines \cite{isensee2021nnunet} and transformer-based
architectures \cite{hatamizadeh2022unetr, hatamizadeh2021swinunetr}
improve accuracy further at higher cost. Conversely,
CASCADE \cite{rahman2023cascade} refines encoder features through a
cascaded attention decoder, and EMCAD \cite{rahman2024emcad} reduces
decoding cost sharply by combining multi-scale depth-wise convolutions
with channel, spatial, and grouped gated attention. EMCAD, however, has been validated only in 2D so far. It is precisely in 3D settings, where
memory and compute constraints are most acute, that such an efficient
design would yield the greatest benefit. Beyond this architectural gap, a
methodological gap exists in the literature: new decoders are typically
introduced alongside different encoders, schedules, and data partitions.
As a result, performance gains cannot be definitively attributed to the
decoder alone and are commonly reported as aggregate scores without
testing whether the observed differences are statistically meaningful.

In this work, we address both gaps. We extend EMCAD to volumetric
segmentation and systematically evaluate it under a controlled
experimental protocol in which the decoder is the only variable, on the
BraTS 2020 benchmark \cite{bakas2018identifying}. Specifically, our
contributions are as follows: (i) we present 3D-EMCAD, a volumetric
formulation of the EMCAD decoder in which all operations are
three-dimensional and normalization is adapted to the small batch sizes
imposed by volumetric training; (ii) we conduct a controlled comparison
of five decoders sharing an identical encoder, schedule, and folds, with
a complete accounting of the parameters and computational cost
attributable to each decoder; and (iii) we show, using a paired
significance test across all $369$ cases, that the five decoders are
statistically indistinguishable in accuracy, while 3D-EMCAD attains the
performance of substantially larger decoders at the lowest parameter and
computational cost.

\begin{figure}[!t]
\centering
\includegraphics[width=0.94\columnwidth]{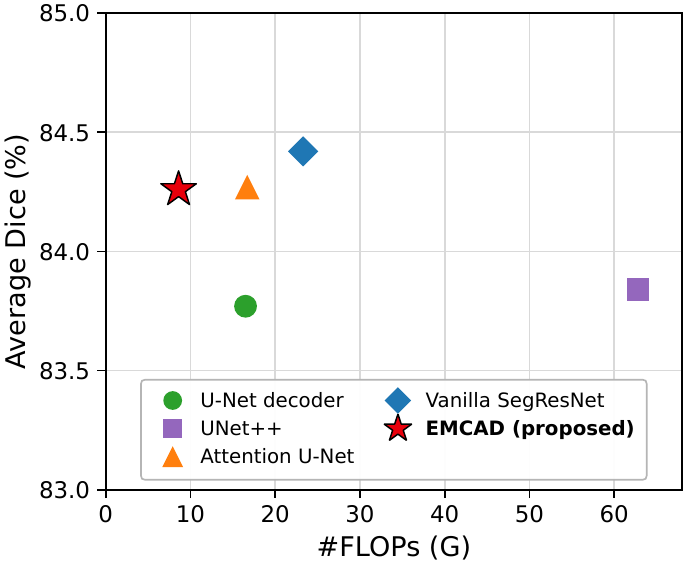}
\caption{Mean Dice versus decoder FLOPs on BraTS 2020. 3D-EMCAD achieves accuracy of the competing decoders with the fewest operations;
UNet++ requires over seven times as many FLOPs without any gain.}
\label{fig:flops}
\end{figure}

\section{Proposed Method}
\label{sec:method}

\begin{figure*}[!t]
\centering
\includegraphics[width=\textwidth]{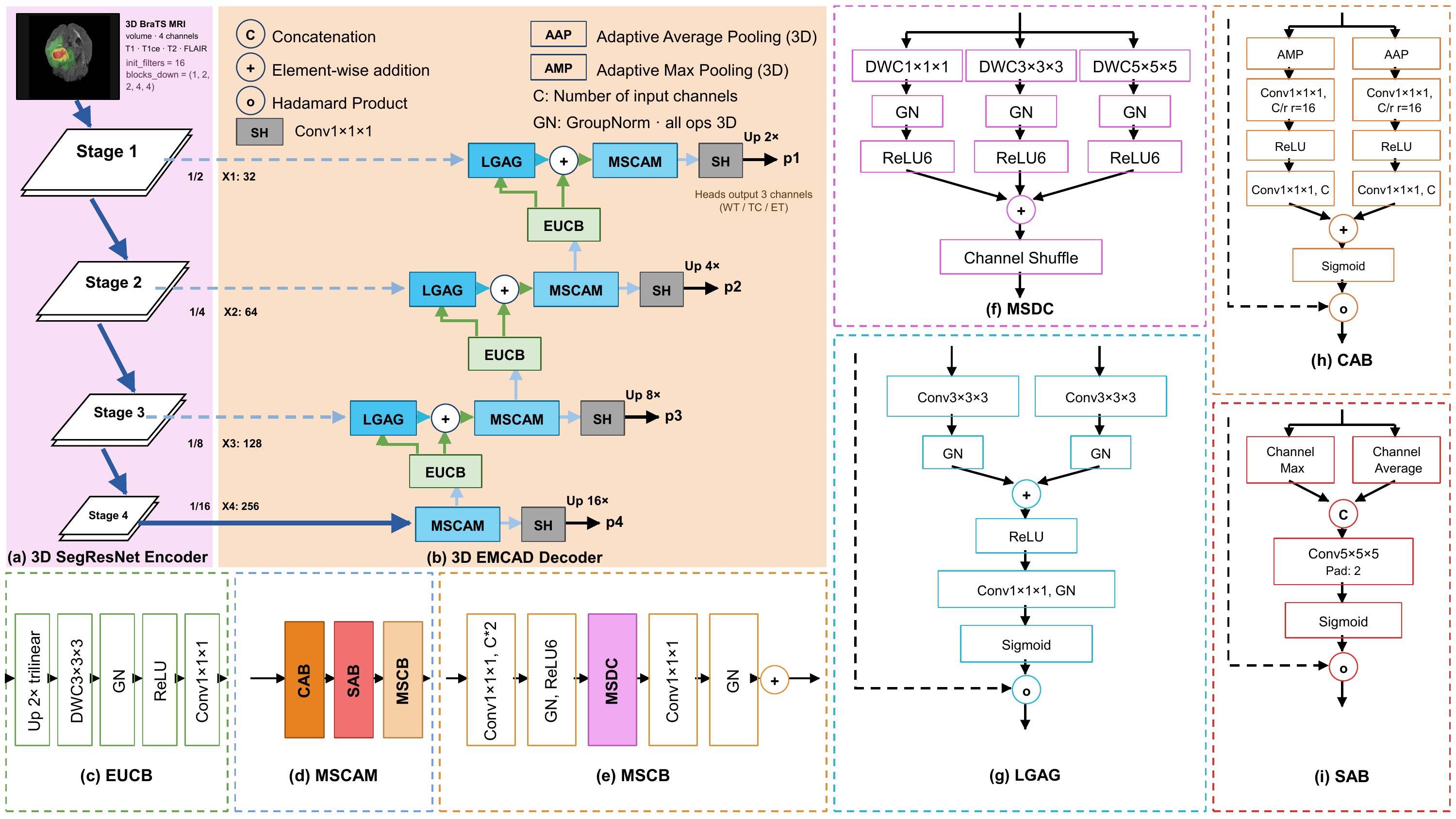}
\caption{The proposed 3D-EMCAD architecture: (a) volumetric SegResNet
encoder producing the pyramid $X_1$--$X_4$; (b) cascaded decoder with
segmentation heads $p_1$--$p_4$; (c) EUCB; (d) MSCAM; (e) MSCB; (f) MSDC;
(g) LGAG; (h) CAB; and (i) SAB. All operations are volumetric, and every
normalization layer is group normalization (GN).}
\label{fig:architecture}
\end{figure*}

This section describes 3D-EMCAD, a volumetric extension of the efficient
multi-scale convolutional attention decoder (EMCAD) proposed for
two-dimensional segmentation \cite{rahman2024emcad}.

\subsection{Overall Architecture}
\label{subsec:overall}
The network follows the encoder--decoder paradigm of
Fig.~\ref{fig:architecture}. A volumetric encoder maps an input volume
$\mathbf{x} \in \mathbb{R}^{C_{\text{in}} \times H \times W \times D}$,
with $C_{\text{in}} = 4$ MRI modalities, to a four-stage feature pyramid
$\{X_1, X_2, X_3, X_4\}$ ordered from the shallowest to the deepest
stage. The decoder is cascaded: the deepest feature map is refined first,
then progressively upsampled, fused with the corresponding skip
connection through a gated attention mechanism, and refined again. It
produces four segmentation maps $\{p_1, \dots, p_4\}$ of $N = 3$ channels,
one per stage and all restored to the input resolution; the finest, $p_1$,
is the final prediction, while the rest provide deep supervision.

Three properties distinguish this formulation from the 2D design: every
operation is volumetric; every batch normalization layer is replaced by
group normalization (GN) \cite{wu2018group}, which is independent of
batch statistics and therefore stable at the small batch sizes volumetric
training requires; and the large-kernel operators are re-sized for 3D
(Section~\ref{subsec:mscam}).

\subsection{Volumetric Encoder}
\label{subsec:encoder}
We adopt a 3D SegResNet backbone \cite{myronenko2019segresnet},
retaining only its contracting path: an initial convolution with
$f = 16$ filters followed by five residual stages of $(1, 2, 2, 4, 4)$
blocks with stride-two downsampling between them. Taking the outputs of
the last four stages,
\begin{equation}
    X_i \in \mathbb{R}^{c_i \times \frac{H}{2^i} \times \frac{W}{2^i}
    \times \frac{D}{2^i}}, \qquad i = 1, 2, 3, 4,
    \label{eq:encoder_features}
\end{equation}
with channel widths $[32, 64, 128, 256]$ and spatial resolutions of
$1/2$, $1/4$, $1/8$, and $1/16$ of the input. The same encoder is used
for every decoder evaluated here, ensuring that the decoder remains the
only variable.

\subsection{Multi-scale Convolutional Attention Module (MSCAM)}
\label{subsec:mscam}
The MSCAM refines each decoder stage by applying in sequence a channel
attention block ($\mathrm{CAB}$), which determines which channels to
emphasize, a spatial attention block ($\mathrm{SAB}$), which determines
where to attend, and a multi-scale convolution block ($\mathrm{MSCB}$),
which enhances features across several receptive field sizes:
\begin{equation}
    \mathrm{MSCAM}(\mathbf{x}) =
    \mathrm{MSCB}\!\left(\mathrm{SAB}\!\left(\mathrm{CAB}
    (\mathbf{x})\right)\right).
    \label{eq:mscam}
\end{equation}

\subsubsection{Channel Attention Block}
The CAB aggregates each channel over the volume by adaptive maximum
pooling $P_m(\cdot)$ and average pooling $P_a(\cdot)$, passes both
descriptors through a shared bottleneck of point-wise convolutions
$C_1(\cdot)$, $C_2(\cdot)$ with reduction ratio $r = 16$ and an
intermediate $\R$, and converts their sum into channel weights via a
sigmoid $\sigma(\cdot)$:
\begin{equation}
    \begin{split}
    \mathbf{w} &= C_2\big(\R(C_1(P_m(\mathbf{x})))\big) \\
               &\quad + C_2\big(\R(C_1(P_a(\mathbf{x})))\big), \\
    \mathrm{CAB}(\mathbf{x}) &= \sigma(\mathbf{w}) \circledast \mathbf{x},
    \end{split}
    \label{eq:cab}
\end{equation}
where $\circledast$ is the Hadamard product; sharing the bottleneck
between both pooling paths halves the cost of this block.

\subsubsection{Spatial Attention Block}
The SAB pools the maximum ($\mathrm{Ch}_{\max}$) and mean
($\mathrm{Ch}_{\text{avg}}$) response along the channel dimension,
concatenates the two single-channel volumes, and applies a large-kernel
convolution $\mathrm{LKC}(\cdot)$ and a sigmoid:
\begin{equation}
    \mathrm{SAB}(\mathbf{x}) = \sigma\!\left(\mathrm{LKC}
    \big([\,\mathrm{Ch}_{\max}(\mathbf{x}),
    \mathrm{Ch}_{\text{avg}}(\mathbf{x})\,]\big)\right)
    \circledast \mathbf{x}.
    \label{eq:sab}
\end{equation}
Since the cost of a volumetric kernel grows cubically rather than
quadratically with its side length, we replace the $7 \times 7$ kernel of
the 2D design by a $5 \times 5 \times 5$ one, preserving a large local
context at substantially lower cost.

\subsubsection{Multi-scale Convolution Block}
The MSCB follows the inverted residual design of
MobileNetV2 \cite{sandler2018mobilenetv2}, replacing the single
depth-wise convolution by a multi-scale one ($\mathrm{MSDC}$). It expands
the channels twofold with a point-wise convolution
$\mathrm{PWC}_1(\cdot)$, GN, and $\Rs$; applies the $\mathrm{MSDC}$;
restores the inter-channel dependencies destroyed by depth-wise filtering
with a channel shuffle $\CS(\cdot)$ \cite{zhang2018shufflenet}; and
projects back with $\mathrm{PWC}_2(\cdot)$ and GN:
\begin{equation}
    \begin{split}
    \mathbf{z} &= \mathrm{MSDC}\!\big(\Rs(\GN(\mathrm{PWC}_1
    (\mathbf{x})))\big), \\
    \mathrm{MSCB}(\mathbf{x}) &= \GN\!\big(\mathrm{PWC}_2
    (\CS(\mathbf{z}))\big).
    \end{split}
    \label{eq:mscb}
\end{equation}
The $\mathrm{MSDC}$ applies a bank of parallel depth-wise convolutions
with different kernel sizes and sums their responses,
\begin{equation}
    \mathrm{MSDC}(\mathbf{x}) = \sum_{k \in \mathcal{K}}
    \Rs\!\big(\GN(\mathrm{DWC}_k(\mathbf{x}))\big),
    \label{eq:msdc}
\end{equation}
where $\mathrm{DWC}_k(\cdot)$ is a depth-wise convolution of kernel size
$k \times k \times k$ and $\mathcal{K} = \{1, 3, 5\}$. This parallel arrangement captures lesion structures of different extents
within a single block, while depth-wise filtering keeps its cost low, the
principal source of the efficiency reported in
Section~\ref{sec:results}.

\subsection{Large-kernel Grouped Attention Gate (LGAG)}
\label{subsec:lgag}
Skip connections transfer high-resolution detail but also carry
background responses irrelevant to the target. The LGAG suppresses these
by gating the skip features with coefficients derived from the coarser
decoder features. Denoting by $\mathbf{g}$ the skip map and $\mathbf{u}$
the upsampled decoder map, both are transformed by separate
$3 \times 3 \times 3$ convolutions $C_g(\cdot)$, $C_u(\cdot)$,
normalized, added, and activated:
\begin{equation}
    q_{\text{att}}(\mathbf{g}, \mathbf{u}) = \R\!\big(
    \GN(C_g(\mathbf{g})) + \GN(C_u(\mathbf{u}))\big),
    \label{eq:lgag_1}
\end{equation}
\begin{equation}
    \mathrm{LGAG}(\mathbf{g}, \mathbf{u}) = \mathbf{g} \circledast
    \sigma\!\big(\GN(C(q_{\text{att}}(\mathbf{g}, \mathbf{u})))\big),
    \label{eq:lgag_2}
\end{equation}
where $C(\cdot)$ reduces the gate to a single channel. Unlike the
point-wise gating of the attention U-Net \cite{oktay2018attention}, the
$3 \times 3 \times 3$ kernels in \eqref{eq:lgag_1} estimate the gate from
a larger neighborhood, which suits the diffuse boundaries of glioma
sub-regions.

\subsection{Efficient Up-convolution Block (EUCB)}
\label{subsec:eucb}
The EUCB doubles the resolution of a feature map and matches its channel
count to the next shallower stage using trilinear upsampling
$\mathrm{Up}(\cdot)$, a $3 \times 3 \times 3$ depth-wise convolution
$\mathrm{DWC}(\cdot)$ with GN and $\R$, and a point-wise convolution
$C_{1 \times 1 \times 1}(\cdot)$:
\begin{equation}
    \mathrm{EUCB}(\mathbf{x}) = C_{1 \times 1 \times 1}\!\big(
    \R(\GN(\mathrm{DWC}(\mathrm{Up}(\mathbf{x}))))\big).
    \label{eq:eucb}
\end{equation}
Depth-wise filtering and interpolation, rather than dense and transposed
convolutions, make this block considerably cheaper than a conventional
upsampling stage.

\subsection{Cascaded Decoding and Segmentation Heads}
\label{subsec:cascade}
Decoding starts at the deepest stage, where $X_4$ is refined directly,
and proceeds towards the input resolution; for $i = 3, 2, 1$ the refined
map $\mathbf{d}_{i+1}$ is upsampled and fused with its skip connection:
\begin{align}
    \mathbf{d}_4 &= \mathrm{MSCAM}_4(X_4), \label{eq:dec_deep}\\
    \mathbf{u}_i &= \mathrm{EUCB}_{i+1}(\mathbf{d}_{i+1}),
    \label{eq:dec_up}\\
    \mathbf{d}_i &= \mathrm{MSCAM}_i\!\big(
    \mathrm{LGAG}_i(X_i, \mathbf{u}_i) + \mathbf{u}_i\big).
    \label{eq:dec_fuse}
\end{align}
The residual form of \eqref{eq:dec_fuse} preserves the upsampled
representation when the gate suppresses the skip connection. Each stage
ends in a segmentation head $\mathrm{SH}_i(\cdot)$, a point-wise
convolution mapping $c_i$ channels to the $N$ regions, followed by
trilinear interpolation $\mathcal{I}(\cdot)$ to the input resolution:
\begin{equation}
    p_i = \mathcal{I}\!\big(\mathrm{SH}_i(\mathbf{d}_i)\big),
    \qquad i = 1, 2, 3, 4.
    \label{eq:seg_head}
\end{equation}
At inference, the three nested regions are obtained from $p_1$ using a
sigmoid and a threshold of $0.5$.

\subsection{Deep Supervision and Training Objective}
\label{subsec:loss}
All four heads are supervised. For a prediction $p$ and ground truth $y$,
we combine a soft Dice term, which addresses the severe
foreground--background imbalance of volumetric tumor data, with a binary
cross-entropy term providing stable voxel-wise gradients:
\begin{equation}
    \mathcal{L}(p, y) = 1 - \frac{1}{N}\sum_{n=1}^{N}
    \frac{2\sum_{v} \hat{p}_{n,v}\, y_{n,v} + \epsilon}
         {\sum_{v} \hat{p}_{n,v} + \sum_{v} y_{n,v} + \epsilon}
    + \mathcal{L}_{\text{BCE}}(p, y),
    \label{eq:loss_single}
\end{equation}
where $\hat{p} = \sigma(p)$, $v$ indexes voxels, $n$ the $N$ regions, and
$\epsilon$ prevents division by zero. The total objective sums the four
head losses,
\begin{equation}
    \mathcal{L}_{\text{total}} = \sum_{i=1}^{4} w_i\,
    \mathcal{L}(p_i, y), \qquad
    w = (1, 0.5, 0.25, 0.125),
    \label{eq:loss_total}
\end{equation}
the weights decreasing geometrically with depth, such that the head
producing the final prediction dominates while the coarser heads act as
auxiliary regularizers.

\section{Experimental Setup}
\label{sec:setup}

\subsection{Dataset and Preprocessing}
\label{subsec:dataset}
We evaluate all models on the BraTS 2020 training
collection \cite{menze2014brats, bakas2017advancing, bakas2018identifying},
comprising $369$ multi-institutional pre-operative glioma cases, each
providing four co-registered, skull-stripped MRI modalities (T1, T1ce,
T2, and T2-FLAIR) at an isotropic resolution of $1\,\text{mm}^3$.
Following the challenge protocol, the labels are converted into the three
nested evaluation regions: enhancing tumor (ET), tumor core (TC), and
whole tumor (WT). Since $\text{ET} \subseteq \text{TC} \subseteq
\text{WT}$, these are represented as three overlapping binary masks,
making the task multi-label rather than multi-class. Preprocessing is
deliberately minimal, ensuring that the comparison reflects the decoders
rather than an elaborate pipeline: each modality is normalized to zero
mean and unit variance over its non-zero voxels, and every volume is cropped to the
bounding box of its non-zero region.

\subsection{Baseline Decoders and Controlled Protocol}
\label{subsec:baselines}
To isolate the contribution of the decoder, we compare 3D-EMCAD against
four established alternatives: the native SegResNet decoder
\cite{myronenko2019segresnet}, which we denote the vanilla decoder; a
volumetric Attention U-Net decoder \cite{oktay2018attention}; a
volumetric UNet++ decoder with dense nested skip pathways
\cite{zhou2018unetpp}; and a plain U-Net decoder
\cite{ronneberger2015unet, cicek20163dunet}. Each decoder is attached to
the identical encoder as in Section~\ref{subsec:encoder}, consumes the same
feature pyramid, and is trained on the same data, folds, schedule, and
loss, with the decoder being the only difference among the five
configurations. Decoders that natively support deep supervision are
supervised using \eqref{eq:loss_total}; the remainder are trained with
the single-head form of \eqref{eq:loss_single}.

\subsection{Implementation Details}
\label{subsec:implementation}
All models are implemented in PyTorch with MONAI \cite{cardoso2022monai}
and trained on a single NVIDIA Tesla T4 GPU with $16$~GB of memory.
Training uses random $96 \times 96 \times 96$ patches with a batch size of
two; to counter the severe foreground imbalance, a patch is centered on a
random whole-tumor voxel with probability $0.7$ and sampled uniformly
otherwise. Augmentation consists of random flips along the three spatial
axes and multiplicative intensity scaling drawn from $[0.9, 1.1]$.
Networks are optimized for $100$ epochs with
AdamW \cite{loshchilov2019adamw} and a weight decay of $10^{-5}$, the
learning rate being warmed up linearly over eight epochs to $10^{-4}$ and
then decayed to $10^{-6}$ on a cosine schedule, with gradients clipped to
a norm of $0.5$ and mixed-precision arithmetic throughout. At inference,
predictions are formed by sliding-window inference over the full volume
($96^3$ window, overlap $0.5$, Gaussian weighting); the logits of the
finest head are passed through a sigmoid and thresholded at $0.5$. We
additionally report results with test-time augmentation (TTA), obtained by
averaging the probability maps over the eight combinations of axis flips.

\subsection{Evaluation Metrics and Statistical Analysis}
\label{subsec:metrics}
Accuracy is measured by the Dice similarity coefficient,
$\mathrm{DSC}(Y, P) = 2\lvert Y \cap P\rvert / (\lvert Y\rvert +
\lvert P\rvert)$, computed separately for the ET, TC, and WT regions of
the ground truth $Y$ and the prediction $P$; we report the three region
scores and their mean. All results use three-fold cross-validation over the $369$ cases, with
folds generated once from a fixed seed and reused for every decoder; we
report the fold mean and standard deviation. Efficiency is quantified by
the parameters and floating-point operations (FLOPs) of the decoder
alone, for single $96^3$ patch.

Because the accuracy differences are small relative to the fold-to-fold
variation, comparing fold means cannot establish whether any decoder
genuinely outperforms another. We therefore apply a paired two-sided
Wilcoxon signed-rank test \cite{wilcoxon1945} to the per-case mean Dice
scores of all $369$ cases, comparing the proposed decoder against each
baseline; the Wilcoxon test is preferred to a paired $t$-test because
per-case Dice scores are bounded and markedly skewed. The four resulting
$p$-values are corrected for multiple testing using the Holm--Bonferroni
procedure \cite{holm1979} and assessed at the $0.05$ level. We report the
median per-case difference alongside each $p$-value, since with a sample
of this size a statistically detectable difference need not be
practically meaningful.

\section{Results and Discussion}
\label{sec:results}

\begin{table*}[!t]
\centering
\caption{Segmentation performance and decoder cost on BraTS 2020 ($369$
cases, three-fold cross-validation). Region scores are fold means; the
Mean column adds the standard deviation across folds. \#P and \#F are the
decoder parameters and FLOPs, excluding the shared encoder, for one
$96^3$ patch. Rows are ordered by decoder size and the best value per
column is bold; per Section~\ref{subsec:significance}, none of these
differences is significant except for the vanilla decoder, which
is negligible in magnitude.}
\label{tab:main_results}
\footnotesize
\setlength{\tabcolsep}{3pt}
\begin{tabular}{l cc cccc cccc}
\toprule
& \multicolumn{2}{c}{Decoder cost}
& \multicolumn{4}{c}{Dice (no TTA)}
& \multicolumn{4}{c}{Dice (with TTA)} \\
\cmidrule(lr){2-3}\cmidrule(lr){4-7}\cmidrule(lr){8-11}
Decoder & \#P (M) & \#F (G)
& TC & WT & ET & Mean
& TC & WT & ET & Mean \\
\midrule
3D-EMCAD (ours)
& \textbf{1.15} & \textbf{8.6}
& 0.8507 & \textbf{0.9027} & 0.7713 & \textbf{0.8416$\pm$0.0098}
& 0.8537 & 0.9047 & 0.7693 & 0.8426$\pm$0.0100 \\
Vanilla SegResNet
& 1.22 & 23.3
& \textbf{0.8511} & 0.9026 & 0.7687 & 0.8408$\pm$0.0145
& \textbf{0.8547} & \textbf{0.9053} & 0.7726 & \textbf{0.8442$\pm$0.0146} \\
U-Net decoder
& 2.09 & 16.5
& 0.8396 & 0.8996 & 0.7643 & 0.8345$\pm$0.0109
& 0.8446 & 0.9017 & 0.7668 & 0.8377$\pm$0.0097 \\
Attention U-Net
& 2.11 & 16.7
& 0.8419 & 0.8999 & \textbf{0.7778} & 0.8399$\pm$0.0056
& 0.8473 & 0.9042 & \textbf{0.7766} & 0.8427$\pm$0.0044 \\
UNet++
& 3.18 & 62.8
& 0.8477 & 0.8969 & 0.7638 & 0.8361$\pm$0.0067
& 0.8522 & 0.8992 & 0.7639 & 0.8384$\pm$0.0088 \\
\bottomrule
\end{tabular}
\end{table*}

\subsection{Segmentation Accuracy}
\label{subsec:accuracy}
Table~\ref{tab:main_results} reports the performance of the five
decoders, and its most striking feature is how little they differ.
Without test-time augmentation, the five mean Dice scores span only
$0.8345$ to $0.8416$, a spread of $0.0071$ that is smaller than the
fold-to-fold standard deviation of most models ($0.0056$ to $0.0145$):
the variation due to the decoder is smaller than that due to the data
partition. The same holds for augmentation, where the means span
$0.8377$ to $0.8442$.

3D-EMCAD attains the highest mean Dice without augmentation ($0.8416$)
and the vanilla decoder the highest with it ($0.8442$), the two differing
by at most $0.0026$, and the per-region breakdown shows no consistent
winner: the best TC score belongs to the vanilla decoder, the best WT
score without augmentation to 3D-EMCAD, and the best ET score to the
Attention U-Net. Augmentation yields a small uniform gain of $0.0010$ to $0.0034$ that does
not alter the ordering; therefore, the unaugmented results are the
primary comparison.

\subsection{Decoder Efficiency}
\label{subsec:efficiency}
\begin{figure}[!t]
\centering
\includegraphics[width=0.94\columnwidth]{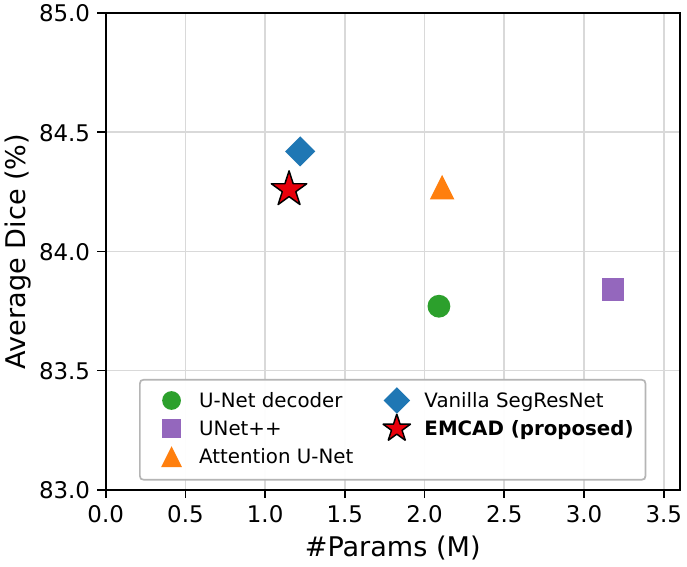}
\caption{Mean Dice versus decoder parameters. The proposed decoder is the
smallest of the five while remaining within the shared accuracy band.}
\label{fig:params}
\end{figure}

Since the decoders are indistinguishable in accuracy, cost becomes the
deciding factor. Fig.~\ref{fig:flops} and Fig.~\ref{fig:params}
plot mean Dice against decoder FLOPs and parameters; in both, the
proposed decoder sits at the extreme left of the accuracy band.

3D-EMCAD requires $1.15$~M parameters and $8.6$~G FLOPs, the smallest of
the five in both respects. The contrast with UNet++ is most pronounced:
at $3.18$~M parameters and $62.8$~G FLOPs, that is $2.8\times$ the
parameters and $7.3\times$ the operations, it attains a mean Dice
$0.0055$ \emph{lower} without augmentation; therefore, its dense nested
skip pathways purchase no accuracy while substantially increasing cost. The
U-Net and Attention U-Net decoders lie between these extremes at roughly
twice the cost, again without an accuracy advantage, and the closest
competitor, the vanilla decoder at $1.22$~M parameters, still needs
$23.3$~G FLOPs, $2.7\times$ those of 3D-EMCAD, using dense rather than
depth-wise convolutions at full resolution. This
efficiency stems from the multi-scale depth-wise convolutions in
\eqref{eq:msdc} and the depth-wise upsampling in \eqref{eq:eucb}, and is
amplified in 3D, where the dense filtering cost grows with $k^3$.

\subsection{Statistical Significance}
\label{subsec:significance}

Since an aggregate comparison of fold means cannot establish whether
these small differences are meaningful, we apply the paired Wilcoxon
signed-rank test of Section~\ref{subsec:metrics} to the per-case mean
Dice scores of across $369$ cases. The test finds no significant difference
between 3D-EMCAD and the UNet++ ($p = 0.38$), Attention U-Net
($p = 0.68$), or U-Net ($p = 0.48$) decoders, all $p$-values being
Holm--Bonferroni corrected. A significant difference arises only relative to the vanilla decoder ($p < 0.001$), with a median per-case margin of
$0.0033$ Dice in its favor. With $369$ paired cases, the test resolves differences of this magnitude,
but three thousandths of a Dice point is negligible in practice and far
below the inter-rater variability reported for glioma
annotation \cite{menze2014brats}. The five decoders are therefore
equivalent in accuracy here, and the appropriate claim is efficiency at
parity rather than superior accuracy.

\subsection{Qualitative Results}
\label{subsec:qualitative}

\begin{figure*}[!t]
\centering
\includegraphics[width=\textwidth]{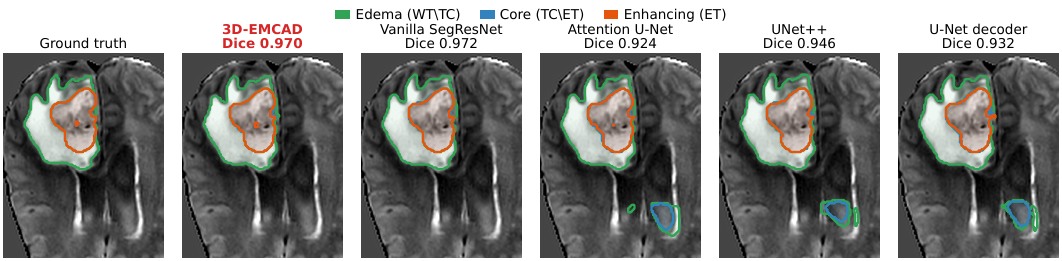}
\caption{Qualitative comparison on a representative validation case; all
panels show the same axial slice with an identical crop and intensity
window.}
\label{fig:qualitative}
\end{figure*}

Fig.~\ref{fig:qualitative} shows the predictions of all five decoders and
the ground truth on an identical axial slice of a representative
validation case. The quantitative parity is evident: every decoder faithfully recovers the extent of the whole tumor and the position of the core, disagreements are confined to the boundaries, particularly to the thin enhancing rim, and no decoder exhibits a failure
mode absent in the others.

\subsection{Discussion and Limitations}
\label{subsec:discussion}
The central observation is that, with the encoder held fixed, the decoder
has remarkably little influence on accuracy: the capacity required to
reconstruct the segmentation from the encoder pyramid appears modest and
is already met by the smallest decoder examined. Decoder capacity is
therefore a reasonable target for reduction when resources are limited.

Three limitations should be acknowledged. First, the evaluation uses a
single dataset; although BraTS 2020 is large and multi-institutional,
generalization to other volumetric targets remains to be verified.
Second, deep supervision is not perfectly controlled, since 3D-EMCAD and
UNet++ are trained with multiple heads, whereas the others are not.
Although this is unlikely to alter the conclusions, given that the
accuracy differences are statistically negligible, it prevents
attributing the residual differences to architecture alone. Third, the
efficiency claim is scoped to parameters and FLOPs: lower FLOPs do not
necessarily translate proportionally into lower latency, since depth-wise
and multi-branch operations map less efficiently onto current GPU kernels
than dense convolutions. Optimizing runtime behavior is left for future
work.

\section{Conclusion}
\label{sec:conclusion}
We have presented 3D-EMCAD, a volumetric extension of the efficient
multi-scale convolutional attention decoder, and evaluated it against
four established decoders on BraTS 2020 under a controlled protocol in
which the encoder, schedule, and cross-validation folds were fixed,
making the decoder the only architectural variable.

Three main conclusions emerge. First, the efficiency of EMCAD transfers
to the volumetric setting: the proposed decoder needs $1.15$~M parameters
and
$8.6$~G FLOPs, the lowest of the five in both respects, and $2.8\times$
fewer parameters and $7.3\times$ fewer operations than UNet++. Second,
this reduction in decoder complexity does not result in a measurable loss
of accuracy: the five decoders differ by at most
$0.0071$ mean Dice, a spread smaller than the fold-to-fold standard
deviation of most of them, and a paired Wilcoxon signed-rank test across
all $369$ cases shows no significant difference between 3D-EMCAD and the
UNet++, Attention U-Net, or U-Net decoders, the only significant
difference amounting to a negligible $0.0033$ median margin against the
vanilla decoder. Third, decoder capacity beyond a modest budget yields no
measurable benefit here, making the decoder a sound target for reduction
when resources are constrained.

The claim established is therefore one of efficiency at parity, scoped to
parameters and operations rather than latency. Future work will examine
whether it holds for other volumetric targets and encoders, and address
the runtime behavior of depth-wise operations on current accelerators.

\balance
\bibliographystyle{IEEEtran}
\bibliography{refs}

\end{document}